\documentstyle[12pt]{article}
\begin{document}
%\draft
%\title{Quark propagator, instantons and gluon propagator}
%\author{L.S. Kisslinger, M Aw, A. Harey and O. Linsuain}}
%\address{Department of Physics, Carnegie Mellon University, 
%Pittsburgh, PA 15213, U.S.A.}
\title{Quark propagator, instantons and gluon propagator}

\author	{L.S. Kisslinger, M. Aw, A. Harey and O. Linsuain\\
	Department of Physics, Carnegie Mellon University, \\
	Pittsburgh, PA 15213, U.S.A.}
 
%\date{\today} 
\maketitle
 
\begin{abstract}
The Schwinger-Dyson formalism is used to check the consistency of instanton
model solutions for the quark propagator with recent models of confining 
gluon propagators. We find that the models are not consistent. A major 
discrepancy is the absence of a vector condensate in the 
instanton model that is present in the solutions with nonperturbative
confining gluons.
\vspace{1cm}

\noindent pacs 12.38.Lg, 12.38.Aw, 13.60.Hb

\end{abstract}
 
%\pacs{ 12.38.Lg, 12.38.Aw, 13.60.Hb}

\newpage 
\section{Introduction}

  The quark propagator contains valuable information
about nonperturbative quantum chromodynamics (NPQCD). In the 
Dyson-Schwinger formalism for QCD  the integral equation for the quark 
propagator involves the dressed gluon propagator and quark-gluon vertex
function, which leads to a set of coupled integral equations.
Recently this formalism has been used to study the quark and gluon
propagators. By using theoretical arguments to express the quark-gluon
vertex in terms of the quark propagator \cite{rw,tan} or other approximations
or models for the dressed vertex function it has been possible 
to constrain the nonperturbative gluon propagator by the   
quark condensates. The quark condensates, which arise from 
chiral symmetry breaking \cite{gor}, have been determined by extensive 
fits to hadronic properties in the QCD sum rule method \cite{svz}.
It was recently shown that space-time structure of the condensates 
\cite{km} as well as the mixed quark condensate \cite{me} can be 
consistently determined in the Dyson-Schwinger form with a confining
gluon propagator.
  
  The nonperturbative quark propagator has also been modeled by
treating quarks propagating in the instanton medium. Instanton models
were used to attempt to understand the various condensates from the
very early days of the QCD sum rule method \cite{svz}. Based on the 
known instanton zero modes \cite{'t}, an instanton medium picture
was shown \cite{sh} to be consistent with the known quark and gluon 
condensates. Subsequently, the propagation of light quarks in the
instanton medium was formulated quantitatively \cite{dp} and refined in 
its treatment of non-zero-mode propagation by Pobylitsa \cite{po}.
Although it does not give rise to a long-range confining force between 
quarks, the instanton vacuum has been shown to provide a good 
phenomenological description of many hadronic properties \cite {sh1}, 
as do QCD sum rules without instantons.  
Instanton effects have been explicitly
included in the nucleon correlator in QCD sum rule calculations of the
nucleon mass \cite{dk,fb} and the nucleon magnetic moments \cite{aw}

  In the present work we use the forms for the nonperturbative quark
propagator derived in the instanton models in the Dyson-Schwinger
formalism and check the consistency of explicit gluonic compared to the 
instanton model treatments of NPQCD.
We avoid the discussion of including instanton
along with other nonperturbative QCD effects, such as condensates and 
Goldstone bosons, without double-counting, by simply seeing if the instanton
model for the quark propagator gives a consistent solution
if one uses the dressed gluon propagators found in the previous Dyson-Schwinger
models within the Dyson-Schwinger framework.

\section{Dyson-Schwinger Form for Quark Propagator}

The quark propagator is defined by
\begin{eqnarray}
 S_q(x) & = & <0|T[q(x)\bar{q}(0)]|0>, 
\label{qp}
\end{eqnarray}
where q(x) is the quark field and T the time-ordering operator. The quark
self-energy, $\Sigma$, is defined by
\begin{equation}
 S_q(p)^{-1}  =  i\hat{p} + m_q + \Sigma(p),
\label{sig}
\end{equation}
with m$_q$ the current quark mass and the notation $\hat{p} = \sum_{\alpha} 
\gamma_{\alpha}p^{\alpha}$, where the $\gamma_{\alpha}$ are Dirac matrices.

Starting with the QCD Lagrangian,
\begin{eqnarray}
\label{qcd}
            L(x) & = & \bar{q}_f(x)(i\hat{D} -m_f)q_f(x) + L^{glue}(x),
\end{eqnarray}
with D$_{\alpha} = \partial_{\alpha}$ -i g$_s$ A$_{\alpha}$(x) and A(x) is
the QCD color field,
one can derive the Dyson-Schwinger equation for the quark 
propagator \cite{iz}.
The self-energy of the quark is given by
\begin{equation}
 \Sigma(p)  =  \int \frac{d^4q}{(2\pi)^4}  g_s ^2 D_{\mu\nu} ^{ab} (p-q) 
\gamma_\mu \frac{\lambda_c ^a}{2} S_q(q)
 \Gamma_\nu^b(q,p),
\label{sd}
\end{equation}
with $D_{\mu\nu} ^{ab} (q)$ the gluon propagator,
$\lambda_c ^a$ the color $SU(3)$ matrix and $\Gamma_\nu^b(q,p)$ the 
quark-gluon
vertex. In the present work we use the approximation $\Gamma_\nu^b(q,p) =
\gamma_\nu \frac{\lambda_c ^b}{2}$, the so-called ``rainbow approximation'',
to be consistent with Ref\cite{km}. It is quite straight-forward for us to
go beyond the rainbow approximation.

Eq. (\ref{sig}) expresses the inverse quark propagator as a perturbative
and a nonperturbative part. Alternatively one can write the propagator as
\begin{eqnarray}
 S_q(x)  & = & S_q^{PT}(x) + S_q^{NP}(x).
\label{qprop}
\end{eqnarray}
For short distances, the operator product expansion for the scalar 
part of $S_q^{NP}(x)$ gives
\begin{eqnarray}
\label{np}
       S_q^{NP}(x) & \simeq &<:\bar{q} (x) q(0): > \\ \nonumber
                   & = & <:\bar{q} (0) q(0): > 
- \frac{x^2}{4} <0|:\bar q (0) \sigma\cdot G(0) q(0) :|0> + \dots ,
\label{ope}
\end{eqnarray}
in which the local operators of the expansion are the quark condensate, 
the mixed condensate, and so forth. G$_{\mu\nu}^c$ is the field tensor
associated with the color field.

We use the Feynman-like gauge and write the gluon propagator as
\begin{equation} 
D^{ab}_{\mu\nu}(q) = \delta^{ab}\delta_{\mu\nu} D(q) .
\label{feynmangauge}
\end{equation}
In the present work we use the results of Ref.\cite{km} for our model
of the function D(q).

An important observation is that the inverse quark propagator can be written
in Euclidean space as
\begin{equation}
     S_q(p)^{-1}  =  i\hat{p} A(p^2) + B(p^2) 
\label{ab}
\end{equation}
Except for the current quark mass and perturbative corrections, the functions
A(p$^2$)-1 and B(p$^2$) are nonperturbative quantities which we refer to as
the vector and scalar propagator condensates, respectively. 
The Dyson-Schwinger
equations (in the Feynman gauge) for A and B are the coupled set
\begin{eqnarray}
[A(p^2) - 1]p^2 & = & \frac{8}{3} g_s ^2 \int \frac{d^4q}{(2\pi)^4} D(p-q)
\frac{A(q^2) }{q^2A^2(q^2) + B^2(q^2)} p\cdot q \nonumber \\
  B(p^2) & = & \frac{16}{3} g_s ^2 \int \frac{d^4q}{(2\pi)^4} D(p-q)
\frac{B(q^2)}{q^2A^2(q^2) + B^2(q^2)} .
\label{eq-sd}
\end{eqnarray}
The quark and mixed quark condensates are given by
\begin{eqnarray}
\label{condensates}
   <:\bar{q}(0) q(0):> & = & -\frac{3}{4\pi^2} \int ds s
  \frac{B(s)}{sA^2(s) + B^2(s)}  \\
   <0|:\bar{q}(0)g \sigma \cdot G(0) q(0):|0> & = & \frac{9}{4 \pi^2}
 \int ds s [ s\frac{B(s)(2-A(s))}{sA^2(s) + B^2(s)} + \nonumber \\
   &  & \frac{81 B(s)[2sA(s)(A(s)-1) + B^2(s)]}{16(sA^2(s) + B^2(s))}],
 \nonumber
\end{eqnarray}
 These equations are all obtained in the rainbow approximation.
The expression for the mixed quark condensate was derived in 
a 1/N$_c$ expansion of the gluon two-point function\cite{me}, making use of 
the self-consistency relations (\ref{eq-sd}). In Ref. \cite{me} the effective
quark mass had reached the current quark mass by 1 GeV, which was taken as
the renormalizaton point and cutoff for the condensate integrals.
We do not use such a cutoff, consistent with Ref. \cite{po}. 
We use the notation  a=-$(2\pi)^2 <:\bar{q}(0)q(0):>$ for the quark
condensate and 
m$_o^2$a = $(2\pi)^2<0|:\bar{q}(0)g\sigma\cdot G(0)q(0):|0>$ for
the mixed quark condensate.
The values of these condensates have been determined by fits to experiment
in QCD sum rule calculations to be a $\simeq$ 0.55 GeV$^3$ and
m$_o^2 \simeq$ 0.8 GeV$^2$, or m$_o^2 a\simeq$ 0.44 GeV$^5$.

The coupled integral equations Eq.(\ref{eq-sd}) have been solved for with
various forms and the parameters fit to the quark condensate. It was shown in
Ref\cite{me} that this gave a satisfactory result for the mixed quark
condensate and in Ref\cite{km} that the nonlocal scale of the nonperturbative
quark propagator as determined from fits to deep inelastic scattering\cite{jk}
was also fit satisfactorily. The range of the confining infra-red gluon
propagator, which is taken as a Gaussian in this work, is very short. In
other calculations\cite{fm} the confining part of the gluon propagator had
a longer range, but the fit to the quark condensate was not very good.
We also use these models for D(s) in our present study.

\section{Instanton Model for Quark Propagator}

In the instanton calculation of Ref \cite{po} the inverse propagator is of the
form (for the current quark mass m$_q$ = 0)
\begin{eqnarray}
\label{po1}
             S_q(p)^{-1} & = & i\hat{p} + B_I(p^2) \nonumber \\
             B_I(p) & = & B_I^0 + O(N/VN_c).
\end{eqnarray}
Using the leading term of a series expansion, the author found the
following instanton solution for the inverse quark propagator, Eq.(\ref{ab}):
\begin{eqnarray}
\label{po2}
            A_I(p) - 1.0 & = & 0.0 \nonumber \\
                 B_I(p) & = & K p^2 f^2(\frac{5}{6} p) \nonumber \\
           f(p) & = & \frac{2}{p} -(3 I_o(p) + I_2(p)) \times K_1(p),
\end{eqnarray}
where $K$ = 0.29 GeV$^{-1}$, and $p$ is in units of GeV. 
We refer to Eq. (\ref{po2}) as the Pobylitsa solution.

An important feature is that the vector propagator condensate
vanishes: A(s) - 1 = 0. This follows from the symmetries of the 
model used by Pobylitsa.

This solution gives an effective quark mass which falls as 1/p$^6$ for
p larger than a few hundred MeV.  Other treatments of the effective quark
mass show different behavior. For example a light-cone model for the pion
form factor \cite{kw} gives B(s) as almost constant and equal to the 
constituent quark mass for p less that about 1 GeV and dropping rapidly
to zero at about 1 GeV. In this model A $\simeq$ 1.0. Let us consider such
a model, which we call the $lc$ model:
\begin{eqnarray}
\label{lc}
           A_{lc}(p) - 1.0 & = & 0.0 \nonumber \\
           B_{lc} & = & \frac{M_N}{3} \ \ {\rm for \ \ p^2 \ \ < \ \ s_o}  \\
                  & = & 0.0 \ \ \ \ {\rm for \ \ p^2 \ \ > \ \ s_o},  \nonumber
\end{eqnarray}
where s$_o$ is a constant to be determined. This $lc$ model can be considered
as a simple alternative to the Pobylitsa model in the sense that the
vector propagator condensate vanishes.
   
   The mixed quark condensate for the instanton model of Ref \cite{dp} was 
estimated\cite{pw} to be m$_o^2$ = 1.4 GeV$^2$, about a factor of
two larger than the empirical QCD sum rule value.

\section{Results and discussion}

  We use for the form of the gluon propagator\cite{fm} $D(s)$
\begin{equation}
{g_s}^2 D(s) \,= \, 
3 \pi^2  \frac{ X^2}{\Delta^2} e^{-\frac{s}{\Delta}}+
 c_u \frac{4\pi^2 d}{s ln(\frac{s}{\Lambda^2}+e)} \, ,
\label{gluon}
\end{equation}
with the parameter c$_u$ = (1.0,0.0) to (include,neglect) the perturbative 
ultra-violet behavior.
The strength parameter X and range parameter $\Delta$ 
are determined by solving
the coupled Dyson-Schwinger equations, Eq.(\ref{eq-sd}), and fitting 
$f_\pi$, the pion decay constant, and the
quark condensate through Eq.(\ref{condensates}). For the Feynman gauge
with c$_u$ = 0.0 these parameters are \cite{km} X = 1.4 GeV and 
$\Delta = 2.0 \times 10^{-3}$ GeV$^2$.

For the $lc$ model, using a = -$(2\pi)^2 <:\bar{q}(0) q(0):> $ = .55 GeV$^3$,
we find s$_o$ = 0.8 GeV$^2$ 
\begin{eqnarray}
\label{i1}
             m_o^2a & = & 0.831  \ \ {\rm GeV^5} \nonumber \\
               B(0) & = & 4.64 \ \ {\rm GeV} \nonumber \\
               A(0) - 1.0 & = & 7.28
\end{eqnarray}
This solution clearly gives inconsistent values for A and B and the mixed
gluon condensate is about a factor of 2 larger than that generally found
in QCD sum rules. Note that this value for the mixed condensate is about
the same as that obtained\cite{pw} with an instanton model\cite{dp}.

For the instanton model, Eq.(\ref{po2}), the value of K found by Pobylitsa
gives the value of B$_I (0) \simeq$ 420 MeV, rather than the constituent 
quark mass. By modifying the factor of K in Eq.(\ref{po2}) by a factor of 
M$_N /(3 \times .420)$ so that B$_I (0)$ = M$_N$/3, we obtain a value of
quark condensate in agreement with the phenomenological one, and
find for the instanton model
\begin{eqnarray}
\label{12}
                     a & = & 0.60 GeV^3 \nonumber \\
               m_o^2 a & = & 2.04  \ \ {\rm GeV^5} \nonumber \\
               B(0) & = & 4.74 \ \ {\rm GeV} \nonumber \\
               A(0) - 1.0 & = & 7.71.
\end{eqnarray}
Therefore we find that the $lc$ and Pobylitsa solutions give similar results
for the Dyson-Schwinger formalism for the quark propagator.
Although the value for the quark condensate agrees with the empirical one, 
the mixed condensate is almost a factor of 5 larger than the empirical
value of .44 GeV$^5$, and about 2.6 times larger than the estimate of 
Ref. \cite{pw}.

The range of the infra-red part of the gluon propagator is very small in
these calculations in which the gluon propagator is constrained by
the phenomenological value of the quark condensate, a=.55 GeV$^3$.
If we relax this constraint we can use the models of Ref. \cite{fm}
in which longer-range infra-red parts of D(s) were used. The results
for the Pobylitsa model for X=1.53 GeV and $\Delta$ = 0.02 GeV$^2$ 
with c$_u$ = 1.0 and $\Lambda$ = 0.2 GeV are
\begin{eqnarray}
\label{i2}
               B(0) & = & 5.92 \ \ {\rm GeV} \nonumber \\
               A(0) - 1.0 & = & 9.89 \, ;
\end{eqnarray}
and with X = 1.65 GeV and $\Delta$ = 0.2 GeV$^2$
\begin{eqnarray}
\label{i3}
               B(0) & = & 1.83 \ \ {\rm GeV} \nonumber \\
               A(0) - 1.0 & = & 3.14.
\end{eqnarray}
Recall that for all of these quark propagator solutions the value of
B(0) is the value of the constituent quark mass, taken as .313 GeV.
From Eq.(\ref{i3}) we note that for the longest range gluonic confining
propagator, with a range of about 0.8 GeV, the instanton
solution is most nearly consistent, but the value of the quark condensate 
is quite small in that model\cite{fm}.

   Our results show that the instanton model is not consistent with the
Dyson-Schwinger model based on the rainbow approximation using a confining
gluon propagator chosen to fit the condensates and other hadronic properties.
One major difference between the NPQCD calculation solving the
model with a confining gluon propagator and the simple substitution of the
instanton model of the quark propagator Ref\cite{po} in the Dyson-Schwinger 
integrals is the vanishing of the vector propagator condensate in the
instanton model. In future work we plan to investigate the inclusion of
instanton effects along with modified confining gluon propagators.

%\acknowledgements
This work was supported in part by the National Science Foundation grant
PHY-9722143.
 
%\begin{references}


\begin{thebibliography}{99}
\bibitem{rw} C.D. Roberts and A.G. Williams, Prog. Part. Nucl. Phys. {\bf 33} 
 (1994) 477, and references therein.
\bibitem{tan} P. Tandy, Prog. Part. Nucl. Phys.  {\bf 39} (1997) 117,
 and references therein. 
\bibitem{gor} M. Gell-Mann, R.J. Oakes and B. Renner, Phys.Rev. {\bf 175}, 
 (1968) 2195).
\bibitem{svz} M. A. Shifman, A. I. Vainshtein and V. I. Zakharov, 
Nucl.Phys. B {\bf 147} (1979) 385; 448.
\bibitem{km} L.S. Kisslinger and T. Meissner, Phys. Rev. {\bf C57}
 (1998) 1528.
\bibitem{me} T. Meissner,  Phys.Lett. B {\bf 405},(1997) 8.
\bibitem{'t}G. 't Hooft, Phys. Rev. Lett {\bf 37} (1976) 8; Phys. Rev. 
{\bf D14} (1976) 3432.
\bibitem{sh} E.V. Shuryak, Nucl. Phys. {\bf B203} (1982) 93, 116, 140;
 Phys. Reports {\bf 115} (1985) 152.
\bibitem{dp} D.I. Dyakonov and V.Yu Petrov, Nucl. Phys. {\bf B245} (1984)
 259; {\bf B272} (1986) 457; Sov. Phys. JETP {\bf 62} (1985) 204.
\bibitem{po} P.V. Pobylitsa, Phys. Lett. {\bf B226} (1989) 387.
\bibitem{sh1} T. Schafer and E.V. Shuryak, Rev. Mod. Phys. {\bf 70} (1998)
 323.
\bibitem{dk} A.E. Dorokhov and N.I. Kochelev, Z. Phys. {\bf C46} (1990) 281.
\bibitem{fb} H. Forkel and M.K. Banerjee, Phys. Rev. Lett. {\bf 71} (1993) 
484.
\bibitem{aw} M. Aw, M.K. Banerjee and H. Forkel, hep-ph/9902458, to be
 published in Phys. Lett. {\bf B} (1999).
\bibitem{iz} C. Itzykson and J-B. Zuber, 
{\it "Quantum Field Theory} (McGraw-Hill Book Co., 1985).
\bibitem{fm}M.R. Frank and T. Meissner, Phys. Rev. {\bf C53} (1996) 2410.  
\bibitem{jk} 
H. Jung and L.S. Kisslinger, Nucl.Phys. A {\bf 586}, 682 (1995).
\bibitem{kw}L.S Kisslinger and S.W. Wang, Nucl. Phys. {\bf B399} (1993) 63.
\bibitem{pw} M.V. Polyakov and C. Weiss, Phys. Lett. {\bf B387} (1996) 841.
%\end{references}
\end{thebibliography}
\end{document}